\documentclass[sigconf]{acmart}
\usepackage{graphicx}
\usepackage{array}
\usepackage{caption}

\AtBeginDocument{%
  }

\copyrightyear{2025}
\acmYear{2025}
\setcopyright{acmlicensed}\acmConference[ASPDAC '25]{30th Asia and South
Pacific Design Automation Conference}{January 20--23, 2025}{Tokyo, Japan}
\acmBooktitle{30th Asia and South Pacific Design Automation Conference
(ASPDAC '25), January 20--23, 2025, Tokyo, Japan}
\acmDOI{10.1145/3658617.3698479}
\acmISBN{979-8-4007-0635-6/25/01}

\acmSubmissionID{3001}

\begin{document}

\title{A 10.60 $\boldsymbol{\mu}$W 150 GOPS Mixed-Bit-Width Sparse CNN Accelerator for Life-Threatening Ventricular Arrhythmia Detection}

\author{Yifan Qin\textsuperscript{1}, Zhenge Jia\textsuperscript{1}, Zheyu Yan\textsuperscript{1}, Jay Mok\textsuperscript{2}, Manto Yung\textsuperscript{2}, Yu Liu\textsuperscript{2}, Xuejiao Liu\textsuperscript{2}, Wujie Wen\textsuperscript{3}, Luhong Liang\textsuperscript{2}, Kwang-Ting Tim Cheng\textsuperscript{2}, Xiaobo Sharon Hu\textsuperscript{1}, Yiyu Shi\textsuperscript{1}}
\affiliation{%
  \institution{\textsuperscript{1}University of Notre Dame, \textsuperscript{2}AI Chip Center for Emerging Smart System,
  \textsuperscript{3}North Carolina State University}
  \country{}
}

\renewcommand{\shortauthors}{Qin et al.}

\begin{abstract}
  This paper proposes an ultra-low power, mixed-bit-width sparse convolutional neural network (CNN) accelerator to accelerate ventricular arrhythmia (VA) detection. The chip achieves 50\% sparsity in a quantized 1D CNN using a sparse processing element (SPE) architecture. Measurement on the prototype chip TSMC 40nm CMOS low-power (LP) process for the VA classification task demonstrates that it consumes 
  10.60 $\mu$W of power while achieving
  a performance of 150 GOPS and a diagnostic accuracy of 99.95\%. The computation power density is only 0.57 $\mu$W/mm$^2$, which is 14.23$\times$ smaller than state-of-the-art works, making it highly suitable for implantable and wearable medical devices.
  \end{abstract}

\maketitle

\vspace{-5pt}
\section{Introduction}

Ventricular fibrillation and ventricular tachycardia are critical types of ventricular arrhythmias (VAs), which are leading causes of sudden cardiac death and significantly contribute to morbidity and mortality. For individuals at high risk of sudden cardiac death, implantable cardioverter-defibrillators (ICDs) serve as a crucial safeguard by providing timely detection and appropriate defibrillation treatment during life-threatening VAs. However, the detection methods employed by current ICDs remain rudimentary, relying on outdated rule-based methods that have not seen significant advancements in decades. While AI-based detection techniques \cite{fan2024ultra,xing202210,zhao201913} have demonstrated considerable promise in the realm of heart disease, a critical challenge that hinders their practical application in embedded implantable devices is the need for real-time response with precise detection in resource-limited platforms. This prompts an urgent need to develop and implement neural network (NN) accelerators that offer low-latency, energy-efficient, and highly accurate inference detection dedicated to these memory and energy-limited platforms.

To this end, we choose a 1D CNN structure due to its excellent feature extraction capability with a relatively small parameter size, which aligns well with our problem scope, and propose a full-stack design that spans the User Interface (UI), compiler, and a fabricated chip. From the algorithmic perspective, our solution employs sparsification and quantization techniques to compress the network. On the hardware side, we introduce a novel sparse processing element (SPE) architecture that is both area- and power-efficient, featuring multi-bit-width configurable multipliers (CMUL). This architecture delivers low latency, reduced energy consumption, and accurate inference. Our design provides a high-accuracy, small-model-size CNN solution for resource-limited medical platforms. These enhancements, achieved through the hardware/software co-design principle, significantly improve energy efficiency and flexibility, enabling the chip to adapt to varying precision and energy consumption requirements in the context of VA detection.
\vspace{-7pt}
\section{Proposed Design}

\begin{figure}[t]
  \centering
  \includegraphics[scale=0.25]{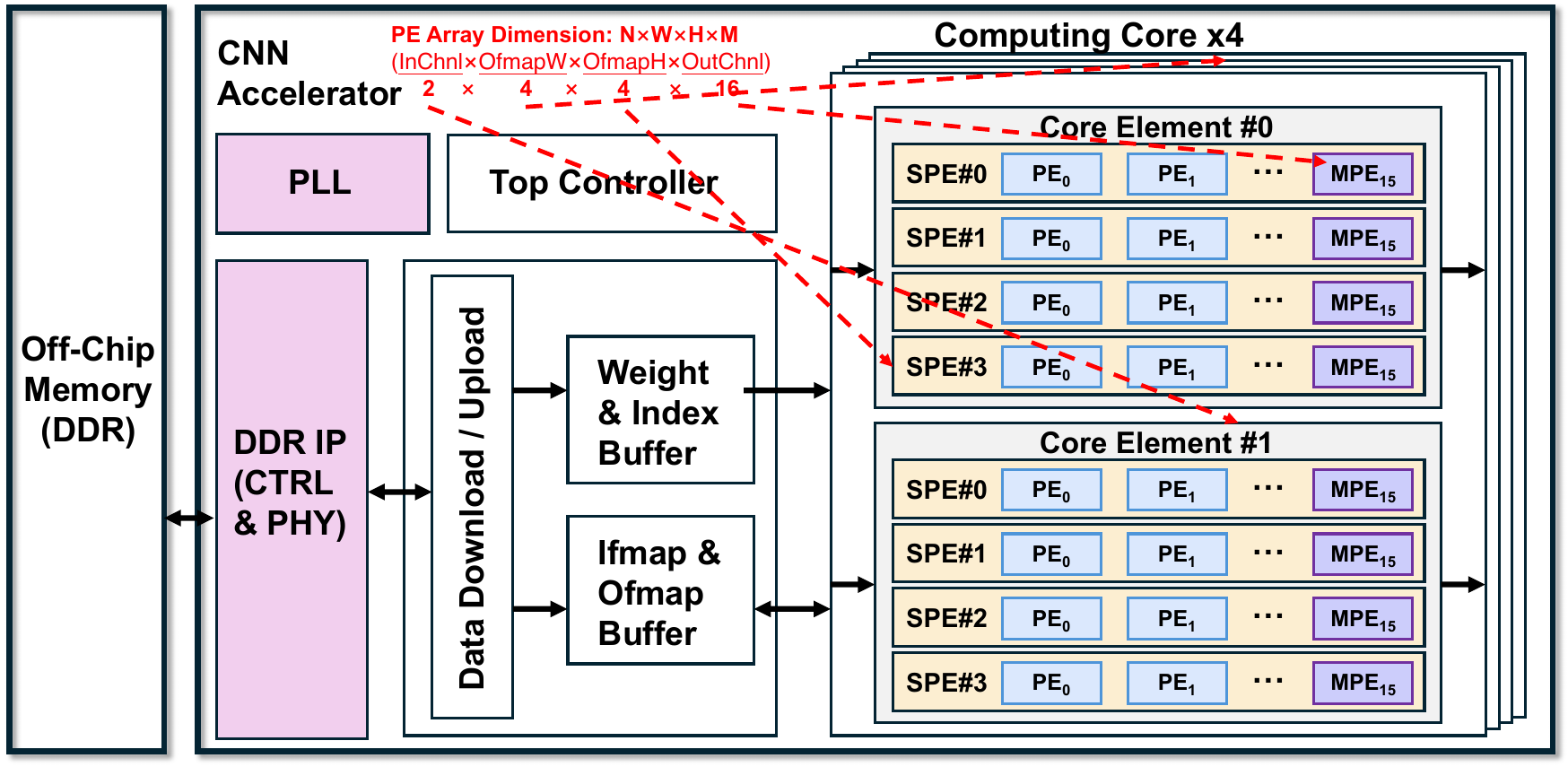}
  \caption{Architecture of the Proposed CNN Accelerator.}
  \Description{Chip arch.}
  \label{fig:chip arch}
  \vspace{-21pt}
\end{figure}

Our NN model is an 8-layer, one-dimensional, fully convolutional network that utilizes co-design pruning with 50\% sparsity and hardware-aware quantization with 8-bit precision. Note that our accelerator also supports mixed precision models and two-dimensional convolutional operation.

Figure \ref{fig:chip arch} shows the architecture of our proposed CNN accelerator. The design features a four-dimensional architecture hierarchically organized into N core elements, W computing cores, H Sparse PEs (SPEs), and M PEs. Here, N, W, H, and M correspond to the input channel number, output feature map width, height, and output channel number, respectively. The accelerator computes a W$\times$H$\times$M block of output feature maps in parallel. We fabricate the accelerator with NxWxHxM as 2x4x4x16 and 12 PEs and 4 Mixed-PEs(MPEs) in 1 SPE, as shown in Figure \ref{fig:chip arch}, redundant computing units will be padded by zero during inference. In this 1d CNN demo, N is padded to 4 and only 1 computing core is used.

\begin{figure}[t]
  \centering
  \includegraphics[scale=0.3]{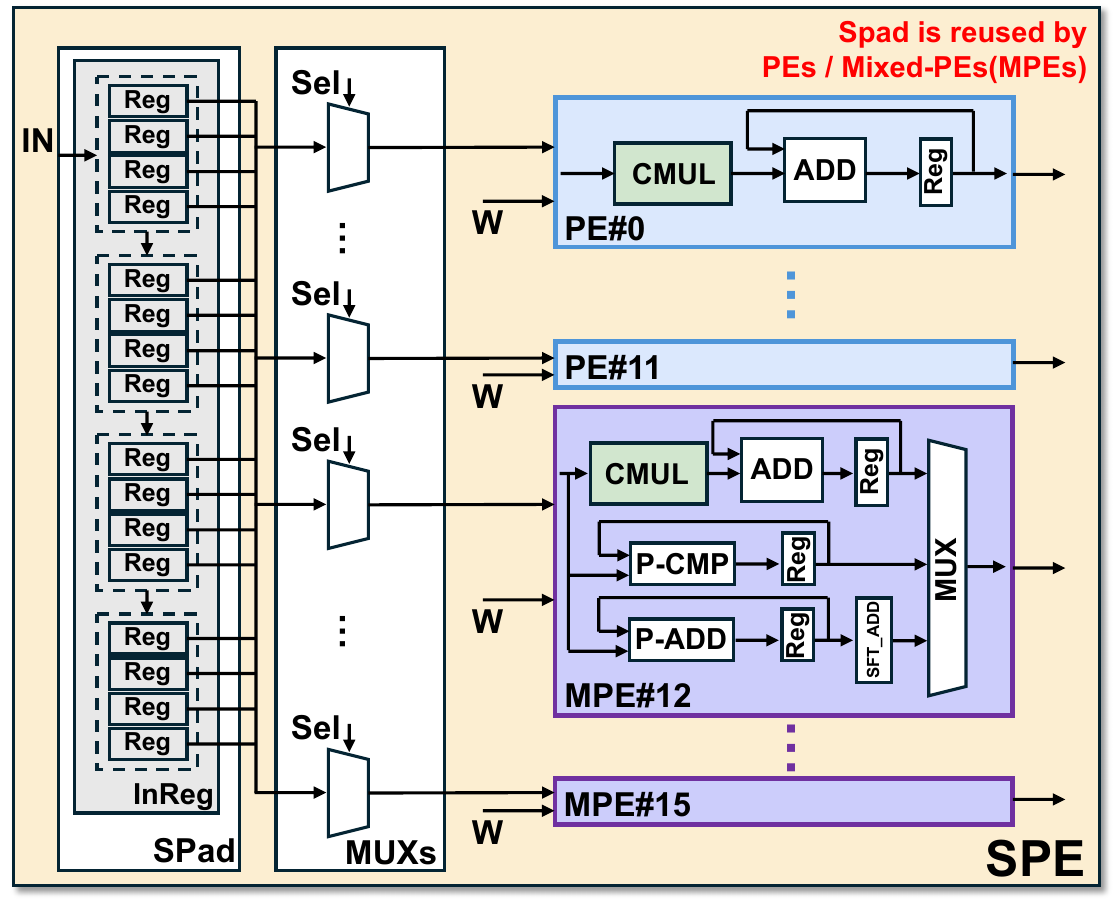}
  \caption{Area-Power-Efficiency SPE Architecture with Single SPad method.}
  \Description{SPE arch.}
  \label{fig:SPE arch}
  \vspace{-15pt}
\end{figure}

\begin{figure}[t]
  \centering
  \includegraphics[scale=0.29]{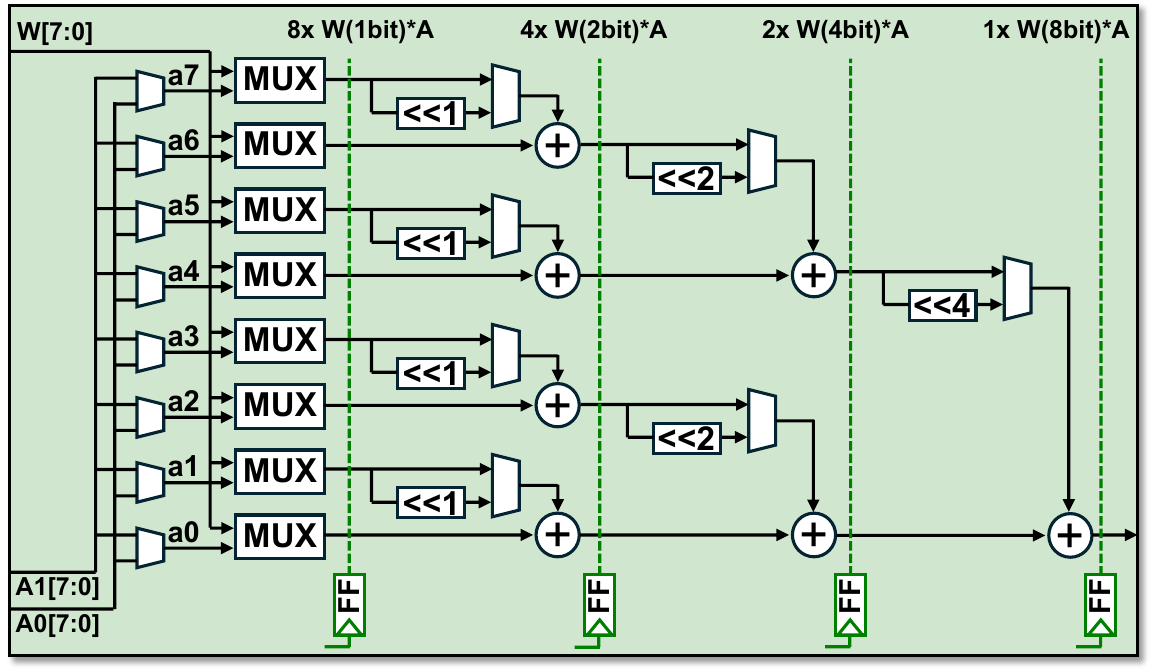}
  \caption{Mixed-bit Signed Reconfigurable Multiplier (CMUL) Architecture.}
  \Description{CMUL arch.}
  \label{fig:CMUL arch}
  \vspace{-23pt}
\end{figure}

Figure \ref{fig:SPE arch} presents the proposed area-power-efficient SPE architecture for sparse CNN acceleration. Each processing element selects input activations from 16 registers, using sparse weights and skipping zero values. The SPE has 12 PEs and 4 Mixed-PEs (MPEs) to process 16 output channels in parallel. Input feature map data is sent to the SPE along the channel direction. Each PE performs multiply-accumulation (MAC) with a configurable bit width, and MPEs additionally support max/average pooling operations. In the previous design \cite{chen2019eyeriss}, multiple scratch pads (SPads) are required since each PE connects to its own local SPad, but in our design, a single SPad module is shared among all PEs and MPEs, allowing them to read inputs and compute simultaneously. This design enables the weight and select signals to be read directly from the on-chip buffers, eliminating the need for FIFO, which consumes significant hardware resources. As a result, only simple control logic is required since all 512 PEs and MPEs operate synchronously, removing the complexity of asynchronous control logic.

Figure \ref{fig:CMUL arch} illustrates the reconfigurable multiplier unit (CMUL) design in PEs and MPEs for mixed-bit signed multiplication. The proposed CMUL supports 8/4/2/1-bit multiplication. In the CMUL, weights are divided into 1-bit segments that multiply corresponding input activation from a multiplexer (MUX). The resulting products are then shifted and accumulated to produce the final output. This design enables CMUL to adaptively select operands for different precision requirements, enhancing both energy efficiency and performance. Furthermore, a co-design pruning mechanism is implemented in the compiler to balance workloads and execution times across and within PEs. The mixed precision and pruning compress weights and activations, minimizing storage and data movement to save energy and computation cycles.

The patient data (provided by SingularMedical) consists of single-lead intracardiac electrograms (IEGM) from the lead RVA-Bi of ICDs. Each recording samples 512 points at a rate of 250 Hz and is pre-processed with a band-pass filter (15-55 Hz). In the demonstration, each IEGM recording is sampled from the patient's ICD and sent to the accelerator chip for detection. The inference results from 6 recordings are aggregated through voting to obtain a diagnosis.
\vspace{-5pt}
\section{Measurement Results}

\begin{figure}[t]
  \centering
  \includegraphics[scale=0.18]{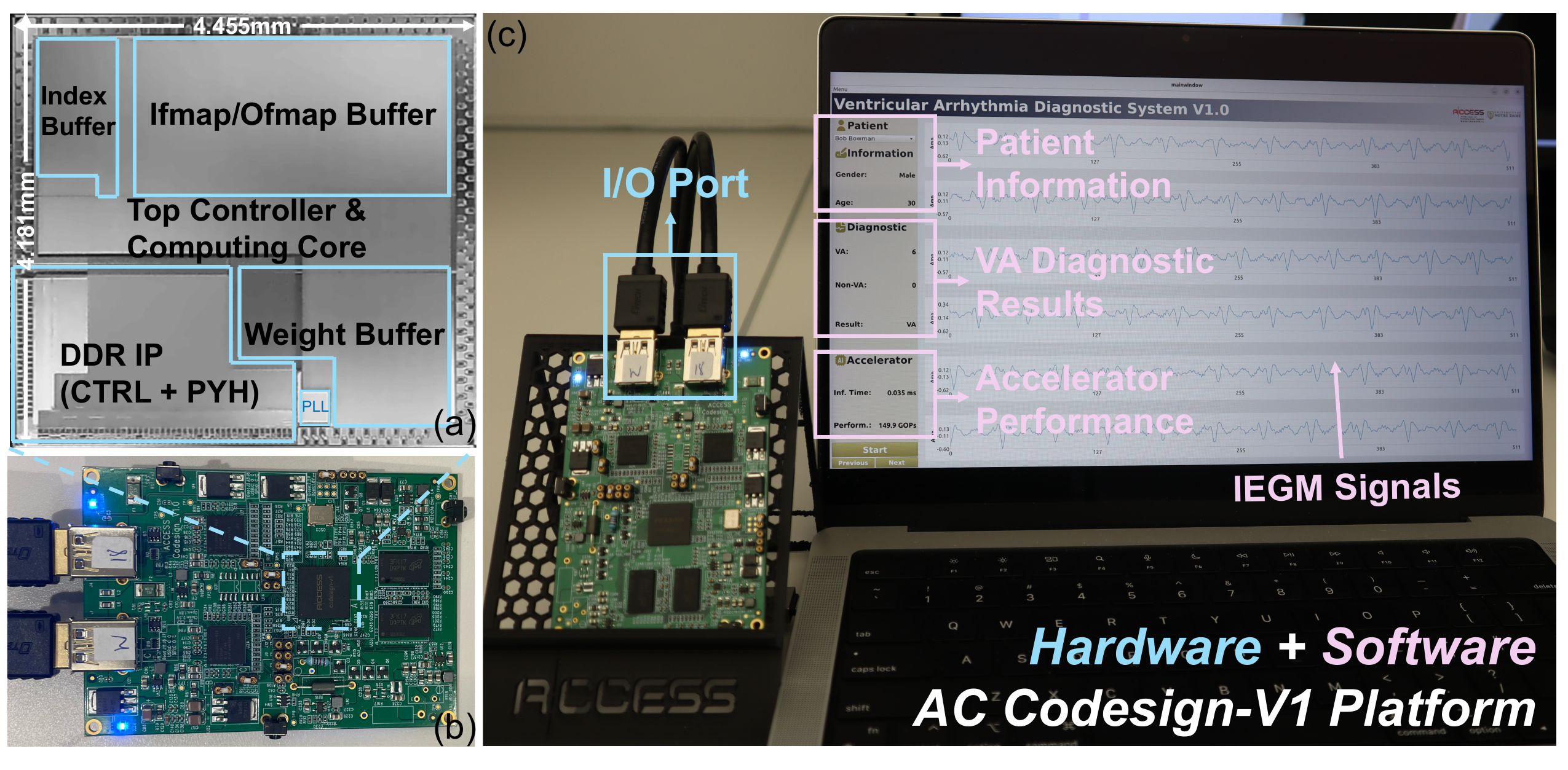}
  \caption{Die Micrograph and the AC Codesign-V1 Platform Demonstration.}
  \Description{Chip Demo.}
  \label{fig:demo}
  \vspace{-10pt}
\end{figure}

\begin{table}[t]
\scriptsize
\caption{Comparison with Previous Works}
\begin{tabular}{|>{\centering\arraybackslash}p{1.2cm}|>{\centering\arraybackslash}p{1cm}|>{\centering\arraybackslash}p{1cm}|>{\centering\arraybackslash}p{1.2cm}|>{\centering\arraybackslash}p{1cm}|c|}
\hline
\textbf{}                                                        & TBCAS'19\cite{zhao201913} & ICICM'22\cite{zhou2022low} & MWSCAS'22\cite{xing202210} & ISCAS'24\cite{fan2024ultra} & Our Work \\ \hline
\begin{tabular}[c]{@{}c@{}}Technology\\ (nm)\end{tabular}        & 180      & 180      & 40        & 40       & 40       \\ \hline
Sparsity                                                         & No       & No       & No        & No       & Yes      \\ \hline
Feature                                                          & ANN      & KS-test  & ANN/SVM   & SNN      & 1D-CNN   \\ \hline
Type                                                             & ASIC     & ASIC     & ASIC      & ASIC     & ASIC     \\ \hline
Area(mm$^2$)                                                     & 0.92     & 1.45     & 0.54      & N/A       & 18.63    \\ \hline
Voltage(V)                                                       & 1.8      & 1.8      & 1.1       & 1.1      & 1.14     \\ \hline
Freq. (Hz)                                                   & 25M      & 0.26K    & 100M      & 1M       & 400M     \\ \hline
Power($\mu$W)                                                    & 13.34    & 11.76    & 5.10      & 12.19    & 10.60    \\ \hline
\begin{tabular}[c]{@{}c@{}}Power Density\\ ($\mu$W/mm$^2$)\end{tabular} & 14.50    & 8.11     & 9.44      & N/A       & 0.57     \\ \hline
\end{tabular}
\label{table:comparison}
\vspace{-12pt}
\end{table}

The proposed CNN accelerator is fabricated using a TSMC 40nm LP CMOS process. Die photo and demo with UI are shown in Figure \ref{fig:demo}. Table \ref{table:comparison} compares the results with previous works. For each IEGM recording, our accelerator achieves an inference time of 35 $\mu$s and a performance of 150 GOPS MAC operations, with an inference accuracy of 92.35\%, diagnostic accuracy of 99.95\%, precision of 99.88\%, recall of 99.84\%, average power consumption of 10.60 $\mu$W. Our design outperforms state-of-the-art (SOTA) works by 14.23 times with 0.57 $\mu$W/mm$^2$ power density. To accommodate other NN models, we fabricate the chip with 512 PEs and a large area (18.63 mm\textsuperscript{2}), but only 128 PEs are engaged in this 1D CNN inference. For implantable or wearable medical applications, the chip size can be scaled down as needed. These results demonstrate that the accelerator is suitable for real-time VA detection in medical applications, offering low latency and power consumption at high accuracy.

\vspace{-7pt}
\begin{acks}
\vspace{-3pt}
This research was supported by ACCESS-AI Chip Center for Emerging Smart Systems, sponsored by InnoHK funding, Hong Kong SAR.
\end{acks}

\vspace{-7pt}
\bibliographystyle{ACM-Reference-Format}
\bibliography{references}


\begin{thebibliography}{5}


\ifx \showCODEN    \undefined \def \showCODEN     #1{\unskip}     \fi
\ifx \showDOI      \undefined \def \showDOI       #1{#1}\fi
\ifx \showISBNx    \undefined \def \showISBNx     #1{\unskip}     \fi
\ifx \showISBNxiii \undefined \def \showISBNxiii  #1{\unskip}     \fi
\ifx \showISSN     \undefined \def \showISSN      #1{\unskip}     \fi
\ifx \showLCCN     \undefined \def \showLCCN      #1{\unskip}     \fi
\ifx \shownote     \undefined \def \shownote      #1{#1}          \fi
\ifx \showarticletitle \undefined \def \showarticletitle #1{#1}   \fi
\ifx \showURL      \undefined \def \showURL       {\relax}        \fi
\providecommand\bibfield[2]{#2}
\providecommand\bibinfo[2]{#2}
\providecommand\natexlab[1]{#1}
\providecommand\showeprint[2][]{arXiv:#2}

\bibitem[Chen et~al\mbox{.}(2019)]%
        {chen2019eyeriss}
\bibfield{author}{\bibinfo{person}{Yu-Hsin Chen}, \bibinfo{person}{Tien-Ju Yang}, \bibinfo{person}{Joel Emer}, {and} \bibinfo{person}{Vivienne Sze}.} \bibinfo{year}{2019}\natexlab{}.
\newblock \showarticletitle{Eyeriss v2: A flexible accelerator for emerging deep neural networks on mobile devices}.
\newblock \bibinfo{journal}{\emph{IEEE Journal on Emerging and Selected Topics in Circuits and Systems}} \bibinfo{volume}{9}, \bibinfo{number}{2} (\bibinfo{year}{2019}), \bibinfo{pages}{292--308}.
\newblock


\bibitem[Fan et~al\mbox{.}(2024)]%
        {fan2024ultra}
\bibfield{author}{\bibinfo{person}{Haodong Fan}, \bibinfo{person}{Liang Chang}, \bibinfo{person}{Junlu Zhou}, \bibinfo{person}{Xi Yang}, \bibinfo{person}{Shuisheng Lin}, {and} \bibinfo{person}{Jun Zhou}.} \bibinfo{year}{2024}\natexlab{}.
\newblock \showarticletitle{An Ultra-Low Power Time-Domain based SNN Processor for ECG Classification}. In \bibinfo{booktitle}{\emph{2024 IEEE International Symposium on Circuits and Systems (ISCAS)}}. IEEE, \bibinfo{pages}{1--5}.
\newblock


\bibitem[Xing et~al\mbox{.}(2022)]%
        {xing202210}
\bibfield{author}{\bibinfo{person}{Rui Xing}, \bibinfo{person}{Li Dong}, \bibinfo{person}{Zhongming Xue}, \bibinfo{person}{Zhuoqi Guo}, \bibinfo{person}{Bingjun Tang}, \bibinfo{person}{Yanze Liu}, {and} \bibinfo{person}{Li Geng}.} \bibinfo{year}{2022}\natexlab{}.
\newblock \showarticletitle{A 10.8 nJ/detection ECG processor based on DWT and SVM for real-time arrhythmia detection}. In \bibinfo{booktitle}{\emph{2022 IEEE 65th International Midwest Symposium on Circuits and Systems (MWSCAS)}}. IEEE, \bibinfo{pages}{1--4}.
\newblock


\bibitem[Zhao et~al\mbox{.}(2019)]%
        {zhao201913}
\bibfield{author}{\bibinfo{person}{Yang Zhao}, \bibinfo{person}{Zhongxia Shang}, {and} \bibinfo{person}{Yong Lian}.} \bibinfo{year}{2019}\natexlab{}.
\newblock \showarticletitle{A 13.34 $\mu$W event-driven patient-specific ANN cardiac arrhythmia classifier for wearable ECG sensors}.
\newblock \bibinfo{journal}{\emph{IEEE transactions on biomedical circuits and systems}} \bibinfo{volume}{14}, \bibinfo{number}{2} (\bibinfo{year}{2019}), \bibinfo{pages}{186--197}.
\newblock


\bibitem[Zhou and Lyu(2022)]%
        {zhou2022low}
\bibfield{author}{\bibinfo{person}{Kanjun Zhou} {and} \bibinfo{person}{Hongming Lyu}.} \bibinfo{year}{2022}\natexlab{}.
\newblock \showarticletitle{A Low-Power Cardiac Signal Processor for Atrial Fibrillation Detection with a Sensitivity of 93.02\%}. In \bibinfo{booktitle}{\emph{2022 7th International Conference on Integrated Circuits and Microsystems (ICICM)}}. IEEE, \bibinfo{pages}{622--625}.
\newblock


\end{thebibliography}

\end{document}